# Magneto-Mechanical Metamaterials with Widely Tunable Mechanical Properties and Acoustic Bandgaps

S. Macrae Montgomery[a§], Shuai Wu[b§], Xiao Kuang[a], Connor Armstrong[a], Cole Zemelka[b], Qiji Ze[b], Rundong Zhang[b], Ruike Zhao[b]*, H. Jerry Qi[a]*

[a]The George W. Woodruff School of Mechanical Engineering, Georgia Institute of Technology, Atlanta, GA 30332, USA

[b]Department of Mechanical and Aerospace Engineering, The Ohio State University, Columbus, OH, 43210, USA

[§] These authors made equal contributions to this work.

*Corresponding authors. Email: zhao.2885@osu.edu; qih@me.gatech.edu;

**Abstract**:

Mechanical metamaterials are architected manmade materials that allow for unique behaviors not observed in nature, making them promising candidates for a wide range of applications. Existing metamaterials lack tunability as their properties can only be changed to a limited extent after the fabrication. In this paper, we present a new magneto-mechanical metamaterial that allows great tunability through a novel concept of deformation mode branching. The architecture of this new metamaterial employs an asymmetric joint design using hard-magnetic soft active materials that permits two distinct actuation modes (bending and folding) under opposite-direction magnetic fields. The subsequent application of mechanical forces leads to the deformation mode branching where the metamaterial architecture transforms into two distinctly different shapes, which exhibit very different deformations and enable great tunability in properties such as mechanical stiffness and acoustic bandgaps. Furthermore, this metamaterial design can be incorporated with magnetic shape memory polymers with global stiffness tunability, which further enables the global shift of the acoustic behaviors. The combination of magnetic and mechanical actuations, as well as shape memory effects, imbue unmatched tunable properties to a new paradigm of metamaterials.

# 1. Introduction

Mechanical metamaterials are architected manmade materials with tailored geometries that exhibit unique behaviors not observed in nature, making them promising candidates for applications such as biomedicine (1), acoustics (2-4), wearable electronics (5, 6), etc. For example, in acoustic applications, the periodic architectures in these metamaterials are designed to have a strong resistance to the propagations of elastic waves within certain frequency ranges (or bandgaps). By properly designing the architectures with targeted bandgaps, these metamaterials enable exciting applications such as acoustic cloaking (7), waveguiding (8-11), and vibration isolation (12, 13). Although the research into mechanical metamaterials is large and rapidly growing, there are still some limitations; a notable one being the lack of tunability. Early designs use static and passive structures and have properties that cannot be changed after the metamaterials are fabricated (3, 14-17). To address this issue, mechanical loads are used to change the metamaterial properties by altering the geometry through buckling or snapping (18-20). At the same time, the recent surge in the interests in smart structures that can dynamically change properties on demand has led to the emergence of active metamaterials (or stimuli-responsive metamaterials), which can transform into predetermined shapes when triggered by certain stimuli. Active metamaterials typically employ stimuli-responsive materials, such as shape memory polymers (SMPs) (21, 22), liquid crystal elastomers (LCEs) (23), and hydrogels (24, 25). These materials can change their shape upon an external stimulus, such as temperature. When incorporated into metamaterials, the structures and their corresponding properties can change in response to stimuli, making them very attractive in applications such as deployable structures (26, 27), soft robotics (28, 29), and tunable acoustic metastructures (23, 30). The tunable properties enabled by stimuli-responsive materials create a wide-open space for new functionality and potential future applications of metamaterials, but at the same time also raises new challenges. For instance, rapid and reversible actuations are highly desirable, but most active materials either cannot be actuated reversibly or have relatively slow response speed. In addition and more importantly, while more versatile properties (or tunability) are preferred, most existing active metamaterials can only achieve two distinct sets of properties. This is because the architectures in most existing metamaterial designs only permit one deformation mode. For example, many active metamaterials use stimuli to trigger snapping for achieving property change where metamaterial structures only have bending deformation(31). In the auxetics metamaterials with chiral structures (3, 32, 33), the joints are designed to only permit rotational motion.

Among the active materials, hard-magnetic soft active materials (hmSAMs) are attractive as they demonstrate rapid and reversible shape changes. These materials consist of a soft elastomer matrix embedded with hard-magnetic particles. Unlike their counterparts (soft-magnetic particles), hard-magnetic particles (or materials) have strong coercivity and can retain strong remanence after magnetization, which generate microtorques under an applied magnetic field. These microtorques drive the hmSAMs to align the programmed magnetization with the applied magnetic field (34), leading to rapid and large shape change. In addition, the desired shape can be programmed during (35-38) or after fabrication (39, 40). Because of these appealing features, hmSAMs have been actively investigated for a variety of potential applications, such as soft robots (39, 41) and biomedical devices (42). The application space of hmSAMs is further expanded by the recent development of magnetic shape memory polymers (M-SMPs) (40), which offer integrated multifunctional shape manipulations including reprogrammable, fast, and reversible shape transformation and locking. M-SMPs can exhibit a modulus change spanning nearly three orders of magnitude as the temperature crosses the glass transition temperature. Such a dramatic change in modulus has been utilized to allow for fast actuation at high temperatures and to hold heavy objects at low temperatures.

In this work, we present a new magneto-mechanical metamaterial that permits great shape and multi-physical property tunability through a novel concept of deformation mode branching, which is achieved by the coupled magnetic actuation and mechanical forces. The architecture of this new metamaterial employs an asymmetric joint design using hmSAMs that allows rapid transition between two distinct actuation modes (bending and folding) under opposite-direction magnetic fields (28). The subsequent application of mechanical forces leads the metamaterial architecture to transform into two distinct shapes, a process referred to as deformation mode branching, which is otherwise unobtainable using either stimulus independently. The deformation mode branching also leads to dramatically different physical behaviors and allows the new metamaterial to have a much broader range of tunability in properties such as mechanical stiffness and acoustic bandgaps. Furthermore, the magneto-mechanical metamaterial design can be incorporated with the M-SMP, which further enhances the shift of the bandgaps with an unprecedented range owing to the dramatic stiffness change. The combination of magnetic and mechanical actuation, as well as the shape memory effects, imbue unmatched tunable properties and can lead to a new paradigm of tunable, functional of metamaterials.

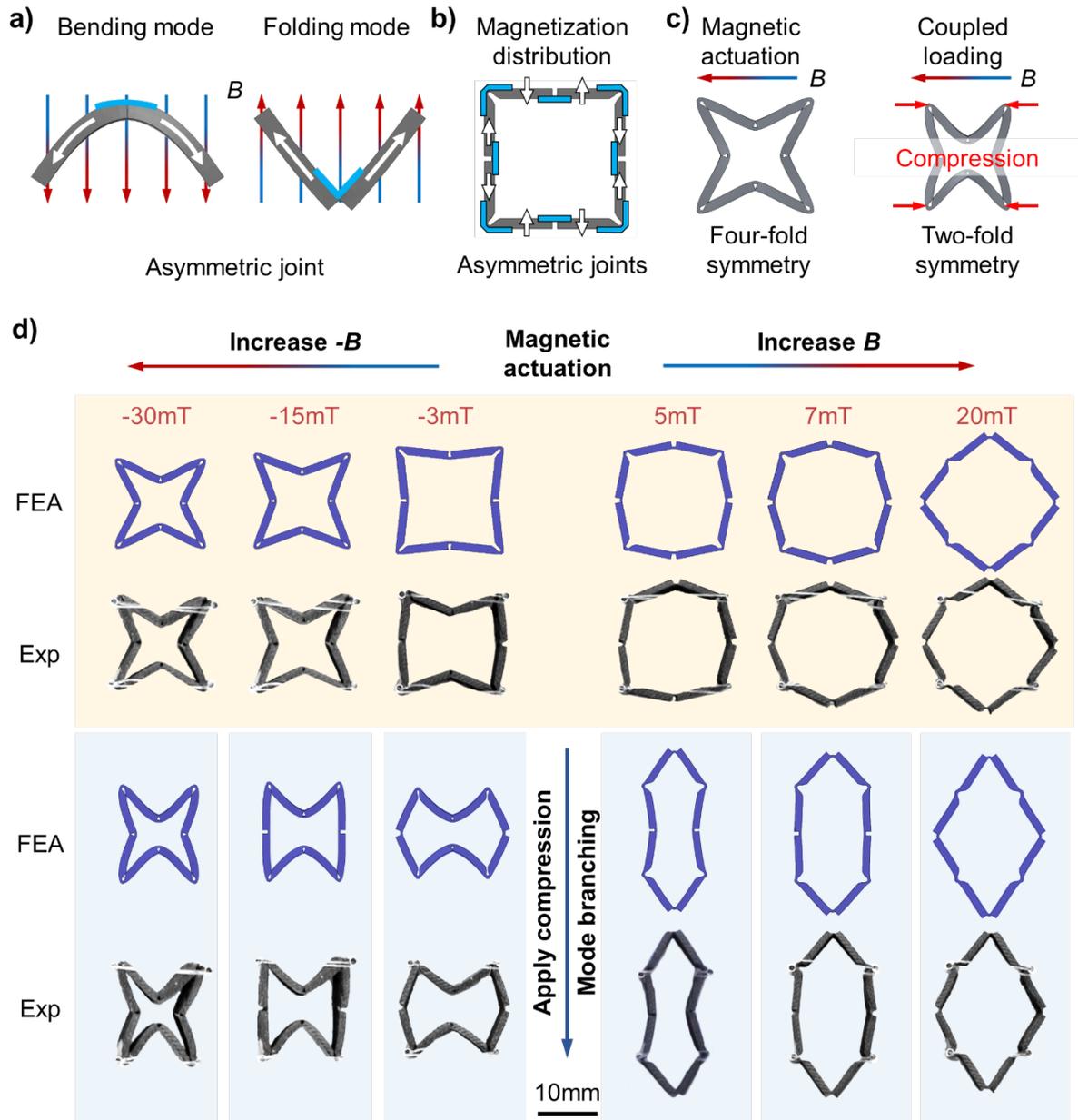

**Figure 1**. The metamaterial unit cell design and the concept of deformation mode branching. a) The actuation of an asymmetric joint showing the folding and bending deformation. b) The schematic of the unit cell with programmed magnetization. c) The unit cell under a magnetic actuation, followed by mechanical compression. d) Deformation mode branching by applying compressions under different magnetic fields shown by both the finite element analysis (FEA) simulations and the experiments.

## 2. Results

### 2.1 Deformation mode branching

The magneto-mechanical metamaterial consists of an array of unit cells. **Figure 1** illustrates the working principle of these unit cells. They are constructed using the concept of the magnetic-responsive asymmetric joint design (28), in which two oppositely polarized hmSAM beams are connected side-by-side with a small gap placed between them. When a downward magnetic field (negative $B$) is applied, the hmSAMs tend to rotate in the opposite directions, closing the gap between them and causing a bending deformation of the joint, referred to as the bending mode (**Figure 1a,** left). As we switch the magnetic field direction to upward (positive $B$), the magnetization directions in the hmSAMs tend to realign with the applied field. Due to the low thickness of the connection region, the joint shows a folding deformation, referred to as the folding mode (Figure 1a, right). The unit cells are then constructed using the asymmetric joints (**Figure 1b**). There are many ways to arrange the asymmetric joints in a unit cell, but we focus this work on illustrating the basic concept of deformation mode branching by using the design shown in Figure 1b. Here, the places of joints on the sidewalls and corners are schematically indicated by the blue highlighted regions. The gaps of the joints on the sidewalls face outward while those at the corners face inward. The magnetization directions of the hmSAMs are also shown in Figure 1b. When a magnetic field $B$ is applied (negative $B$ in **Figure 1c** as an example), the alignment of the magnetization directions of the hmSAMs drives the unit cell to change into shape with a four-fold symmetry (Figure 1c, left). At a predefined applied magnetic field $B$, we hold the magnetic field and apply a compressive force at the unit cell's corners, which breaks the four-fold symmetry down to a two-fold one (Figure 1c, right). This symmetry-reducing mechanism combined with the asymmetric joints achieves the deformation mode branching.

**Figure 1d** shows the deformation mode branching with dramatic shape change under mechanical loading at different magnetic fields, as demonstrated by both the finite element analysis (FEA) simulations and the experiments. The unit cells are fabricated by casting and curing the mixture of hard-magnetic particles (15% vol) and the elastomer resin in 3D printed molds. For each unit cell, four hollow pins are glued to the corners. Four strings are wound through these pins and connected to a fixed plate on one side and to the grip of a universal testing machine on the other. The unit cell is placed inside a pair of Helmholtz coils, which provide a uniform magnetic field along the coils' axial direction (**Figure S1**). During testing the magnetic field is ramped up to the desired value over ten seconds, and the strings are then pulled by the universal testing machine to apply a compressive load. The FEA simulations are conducted by using the commercial FEA software package ABAQUS/Standard (Dassault Systèmes, Providence, RI, USA) in combination with a user element subroutine for hmSAMs (43). See the Experimental Section and Supplementary Information (SI) for details.

As shown in the top left row of Figure 1d, when applying negative magnetic fields, the asymmetric joints exhibit bending mode deformation, causing the unit cell to gradually transform into star-shape with four-fold symmetry. By holding the magnetic field at different intensities (-3mT, -15mT, -30mT) and applying mechanical compression (27%, 22%, 21%, respectively) to the unit cell, the deformation transforms from the four-fold-symmetry shapes to different two-fold-symmetry shapes. The three separate mode branching deformations have distinct features: the -3mT magnetic actuation with mechanical loading leads to shape with two concave sidewalls by bending deformation (top and bottom) and two convex sidewalls by folding deformation (left and right), the -15mT case has shape with two concave sidewalls by bending deformation (top and bottom) and two straight sidewalls (left and right), and the -30mT case leads to a shape with all four concave sidewalls by bending deformation. Under the positive magnetic fields, the asymmetric joints exhibit folding mode deformation, and the shape transforms into a polygon with four-fold symmetry, as shown in the top-right row of Figure 1d. As before, by holding the magnetic field at different intensities (5mT, 7mT, 20mT) and applying compression (31%, 20%, 22%, respectively), mode branching with two-fold symmetry will be achieved. In the three mode branching deformations, the 5mT magnetic actuation with compression leads to shape with two concave sidewalls by bending deformation (left and right) and two convex sidewalls by folding deformation (top and bottom), the 7mT case has shape with two convex sidewalls by folding deformation (top and bottom) and two straight sidewalls (left and right), and the 20mT case leads to a shape with all four convex sidewalls by folding deformation. **Video S1** in the SI shows the transformation of the unit cells.

From the above study of unit cell cases, we can see two distinctly different deformation mode transformations imposed by the mechanical forces: the transformation from the four-fold symmetry to the two-fold symmetry and the transformation of the sidewalls from folding mode to bending mode (or *vice versa*). With these dramatically different shape changes, one can anticipate the deformation mode branching behavior by controlling the coupling of magnetic actuation and mechanical loading, which promotes tunable physical properties due to the diverse structural changes and the introduction of anisotropy to the system. Figure 1d also shows consistency between the FEA simulations and the experimental results, emphasizing the high fidelity of the FEA simulations, which will be used in the study of arrays and bandgap responses.

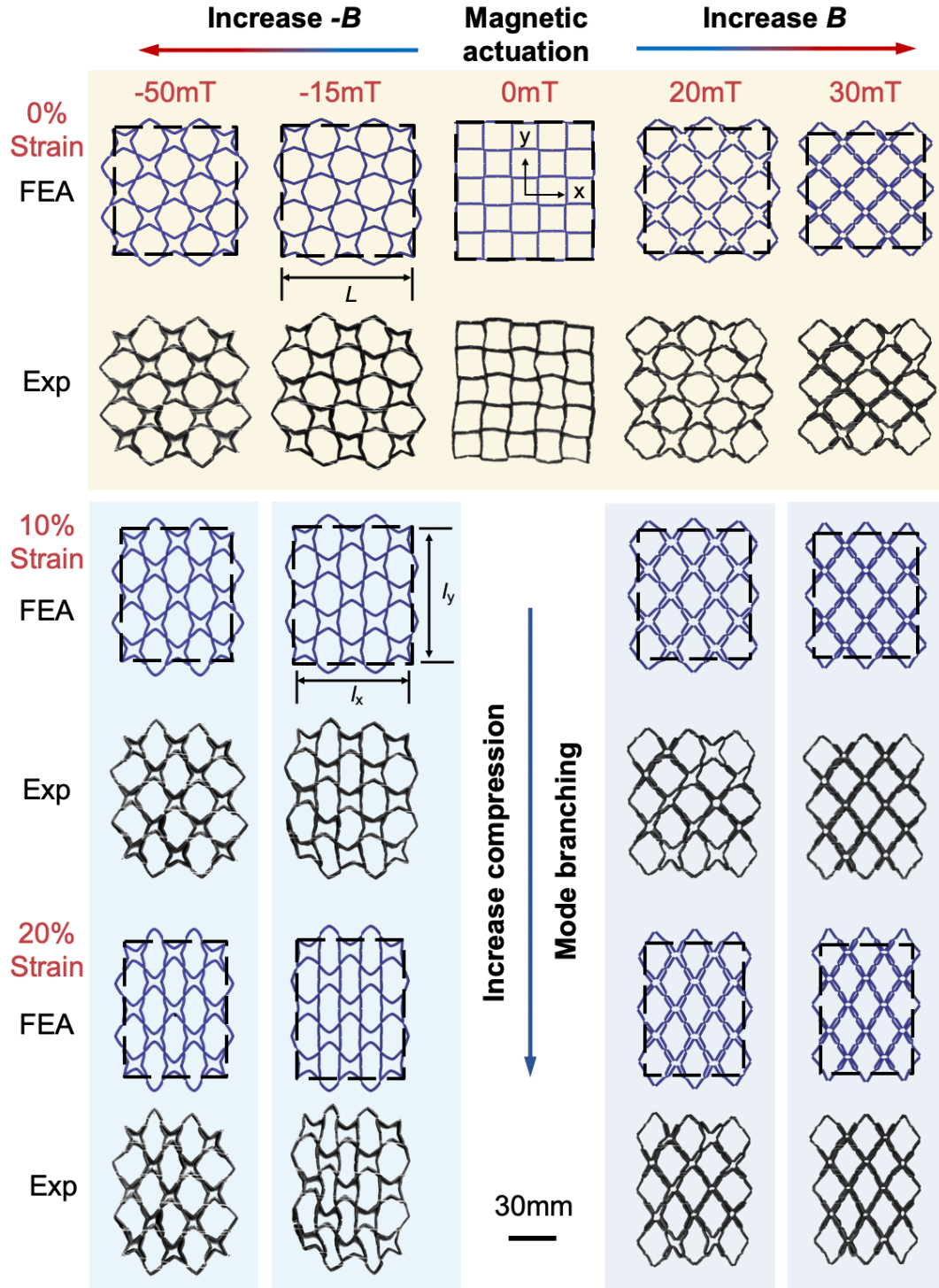

**Figure 2**. Deformation mode branching of a metamaterial array. The top two rows show the shape change of the array under the magnetic fields. The four columns underneath show the shape when the mechanical force is applied under the corresponding magnetic fields. These results indicate the extent of deformation mode branching.

## 2.2 Deformation mode branching in metamaterial array

The previous section discussed the unique shape transformation of the unit cells with the deformation mode branching. In this section, we investigate the deformation mode branching in a metamaterial array constructed by such unit cells. **Figure 2** shows the deformation mode branching of the metamaterial from both the FEA simulations and the experiments. Because the sidewalls of the unit cells would interact with each other to prevent outward deformation if they were side-by-side, we arrange the unit cells in a checkerboard manner. The top-center images in Figure 2 show the as-fabricated metamaterial array (second row) and the FEA model (first row). See the Experimental Section and SI for details regarding sample preparation and the FEA model. The left two columns show the shape change under the negatively applied magnetic fields (-15mT and -50mT) with 0%, 10%, and 20% compressive strains (applied horizontally), respectively. The right two columns show the shape change under the positively applied magnetic fields (20mT and 30mT) with 0%, 10%, and 20% compressive strains, respectively. Since the unit cells are thin-walled structures, the structure contains many openings (or pores). In general, when a magnetic field is applied, these pores have two different deformations: one whose area increases then maintains almost constant during both magnetic and mechanical actuation and one whose area decreases. When a negative magnetic field is applied (bending), the pores with the decreasing area become smaller but cannot be completely closed, owing to the stiffness of the asymmetric joints in the bending mode. In contrast, under a positive magnetic field (folding), the pores with the decreasing area can close completely as the magnetic field or the applied mechanical force increases, leading to a topological change of the metamaterial. For example, under a 30mT magnetic field, these pores disappear and the number of pores is reduced from 25 to 13. The same topological change can be obtained by the compressive force at a small magnetic field. At 20mT magnetic field, applying a 10% compression can eliminate nearly all the small pores. This topological transformation property is, to our best knowledge, not seen in previous works. Therefore, the deformation mode branching introduces two types of deformation modes, one with topological change and one without. It also should be noted that the metamaterial demonstrates slightly different deformation mode branching than the unit cell. This is because of a change in boundary conditions due to the connectivity of neighboring cells. **Video S2** in the SI shows the rapid and reversible magnetic actuation and the coupled magneto-mechanical actuation (at a slower speed to accommodate mechanical compression) of the metamaterial.

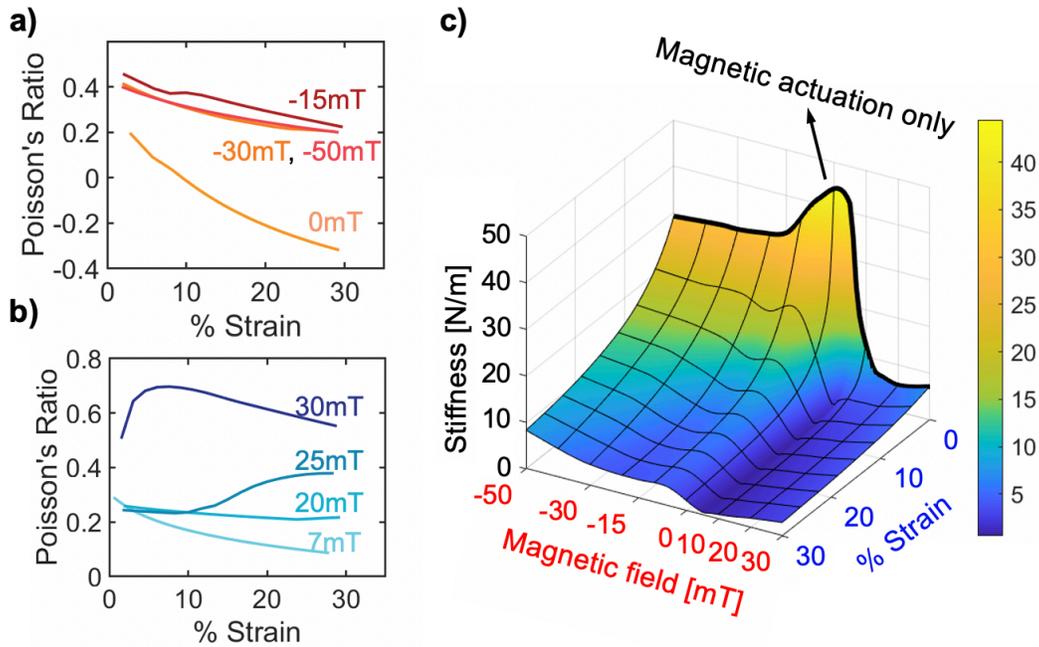

**Figure 3**. Changes in the mechanical properties of the metamaterial during the deformation mode branching. a) The effects of the compressive strain on the Poisson's ratio under negative magnetic fields. b) The effects of the compressive strain on the Poisson's ratio under positive fields. c) 3D contour plot of the metamaterial stiffness as a function of the applied magnetic field and the applied compressive strain. The small-strain stiffness under a pure magnetic field is represented by the bold line.

The deformation mode branching induces dramatic changes in both the local and global architectures of the metamaterial and also leads to mechanical property shifts, as shown in **Figure 3**. The dependence of the Poisson's ratio on both the compressive strain and the applied magnetic field is depicted in **Figures 3a and 3b**. Without a magnetic field (Figure 3a, 0mT), the Poisson's ratio decreases quickly with an increasing strain and becomes negative (auxetic) as the strain passes ~9%. Applying a magnetic field changes the dependence of the Poisson's ratio on the mechanical loading dramatically. When a negative magnetic field is applied, the asymmetric joints deform into a bending mode, which causes the Poisson's ratio to become larger but show less dependence on the magnetic field. For example, under -30mT and -50mT, the Poisson's ratios are almost identical. For -15mT, there is a small kink at the strain around 10%. This kink is due to some horizontal joints in the array changing from bending mode to folding mode (as discussed above) under the applied compressive strain. However, such a transition does not change the Poisson's ratio significantly. In contrast, the positively applied magnetic field can affect the Poisson's ratio substantially (Figure 3b). As discussed previously a positively applied magnetic field leads the array to exhibit the folding mode. Under a small magnetic field, the Poisson's ratio shows less dependence on the compressive load. In particular, at 20mT, the Poisson's is almost a constant (at ~0.22). As the magnetic field increases, as shown in Figure 2, a topology change is observed in the array as the number of the pores reduced. This leads to a new deformation mode, which is only reached at larger strains when the field is low. Larger magnetic fields, on the other hand, lead to a greater initial reduction in pore area, allowing the mode transition to occur at a lower strain (less compression is required to close the pores). This explains why the 25mT case transitions around 15% strain while the 30mT case transitions almost immediately.

**Figure 3c** shows the stiffness contour plots of the metamaterial. Here, the stiffness is calculated by applying a small perturbation mechanical load to the deformed array (under the combined loading of the magnetic field and the mechanical load). The deformation mode branching spawns a diverse stiffness landscape for the structure. At zero strain, where the deformation is induced by the magnetic field only, the stiffness of the metamaterial decreases under both the positively and negatively applied magnetic fields. A much faster decrease is observed under the positive field as the folding mode has a smaller resistance to deformation. Under the negative magnetic field, the stiffness starts to plateau at -15mT when all the asymmetric joints exhibit contact and increase their resistance to deformation. From this point, the geometry does not change significantly as the magnitude of the magnetic field increases. As we apply the mechanical compression, the stiffness is further reduced but more so under the low negative fields than the high negative fields despite their initial similarity. This is because the higher magnetic field can resist the outward folding of the vertical walls which typically leads to a reduced stiffness. The observed global stiffness decrease is similar to post-buckling in thin beam structures where the buckled structure shows a reduction in bending stiffness. Note that in our current work there is no apparent buckling, but the behaviors of the metamaterial show a similarity to those in post-buckling. This is also the reason we name the observed behavior as the deformation mode branching.

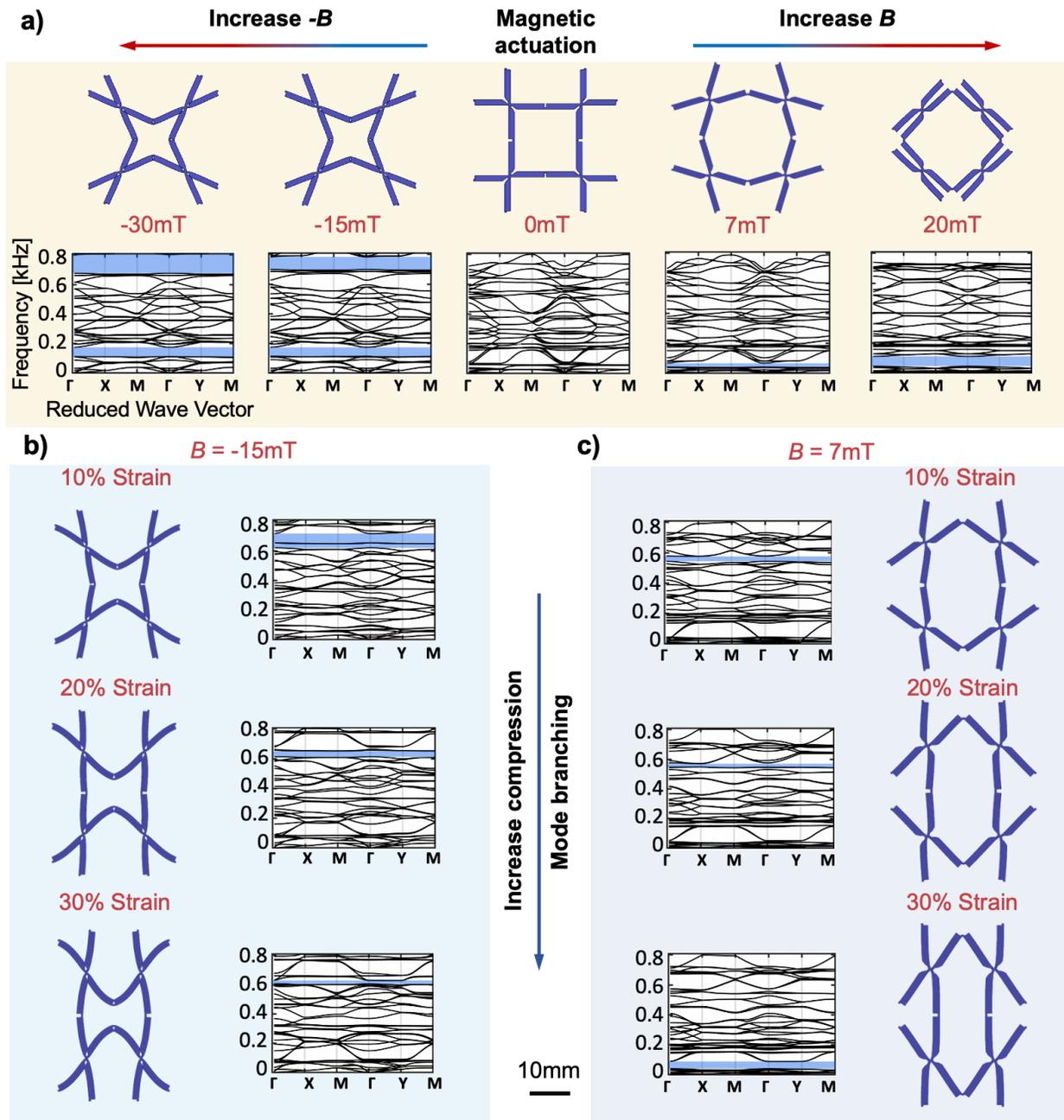

**Figure 4**. Effects of the deformation mode branching to the acoustic bandgaps of the metamaterial. a) The shapes of the representative volume elements (RVEs) under a magnetic field and the corresponding bandgaps shown by Bloch eigenmodes. b) The shapes and the bandgaps of RVEs b) under the negative magnetic field ($B$ = -15mT) and the mechanical compression, and c) under the positive magnetic field ($B$ = 7mT) and the mechanical compression.

**2.3 Tunable acoustic bandgaps due to the deformation mode branching**

We further investigate the metamaterial with the deformation mode branching for acoustic bandgap applications. In metamaterials with periodic structures, self-interference from the deformation of neighboring unit structures prevents elastic waves from propagating within certain frequency ranges, which are known as bandgaps. As discussed above, the coupled magnetic and mechanical loadings lead to dramatic shape changes of our metamaterial array due to the deformation mode branching. This, in turn, leads to dramatic changes in the bandgap response. **Figure 4** shows the bandgap shift enabled by the deformation mode branching. The bandgaps are calculated by using the Bloch wave analysis method (44) with the FEA simulations on the representative volume element (RVE) of the metamaterials. Here, the RVE represents the periodic feature of the array (the top-center of Figure 4), which has a checkboard arrangement and is thus different from the unit cell shown in Figure 1. In Figure 4, the shapes of the RVE under magnetic or magneto-mechanical load are shown alongside plots of the frequencies of their corresponding Bloch eigenmodes with the bandgaps highlighted. **Figure 4a** demonstrates how just an applied magnetic field can shift the bandgaps. The undeformed metamaterial (in the center) exhibits no noticeable bandgaps. As a magnetic field is applied, however, vastly different bandgaps emerge depending on the direction of the magnetic field. Under the negative magnetic field, two bandgaps emerge. At -15mT there are two bandgaps: 115Hz-175Hz, and 699Hz-783Hz. As the magnetic field increases to -30mT, the shape of the RVE does not change much; therefore, the bandgap remains similar to the high-frequency bandgap becoming slightly wider. In contrast, under the positive magnetic field, the 7mT field only introduces a narrow bandgap (42Hz-68Hz) in the low-frequency region. As the magnetic field increases to 20mT, this bandgap becomes much wider (47Hz-112Hz). In both cases, there is no bandgap in the high-frequency region. While the opposite magnetic fields can cause branching into two modes and lead to much different acoustic bandgap properties, variation in the magnetic field alone still provides only limited tunability. However, the magneto-mechanical coupling can significantly broaden the tunability.

**Figures 4b and 4c** show the shift of the bandgaps when a mechanical load is applied. For the case with -15mT magnetic field (Figure 4b), when the mechanical compression is small (10% strain case), the high-frequency bandgap remains relatively unchanged in width but shifts downwards (620Hz-716Hz). However, the change from the four-fold to the two-fold symmetry causes the low-frequency bandgap to quickly vanish. As discussed above, when the compressive strain increases, the sidewalls of the unit cell transition from inward bending to a more vertical position, which causes the high-frequency bandgap to become narrow (614Hz-678Hz at 29%, and 605Hz-628Hz at 30%). Figure 4c shows the periodic cell deformation and the bandgap shift due to the compression under 7mT. Here we can see a very different behavior than the bending case. The mechanical compression causes the low-frequency bandgap to disappear by 10% strain like before, but a new high-frequency bandgap appears between 543Hz-576Hz. This thin bandgap also wanes as the compression is increased to 20% strain. This might be due to the vertical walls beginning to transition from outward folding to a straighter position. At 30% those vertical walls have transitioned into the inwardly bent state and this deformation branching causes the high-frequency gap to complete its closure while a low-frequency gap opens once again around 41Hz-89Hz. These results demonstrate the great tunability enabled by deformation mode branching to significantly change the acoustic response of the metamaterial structure.

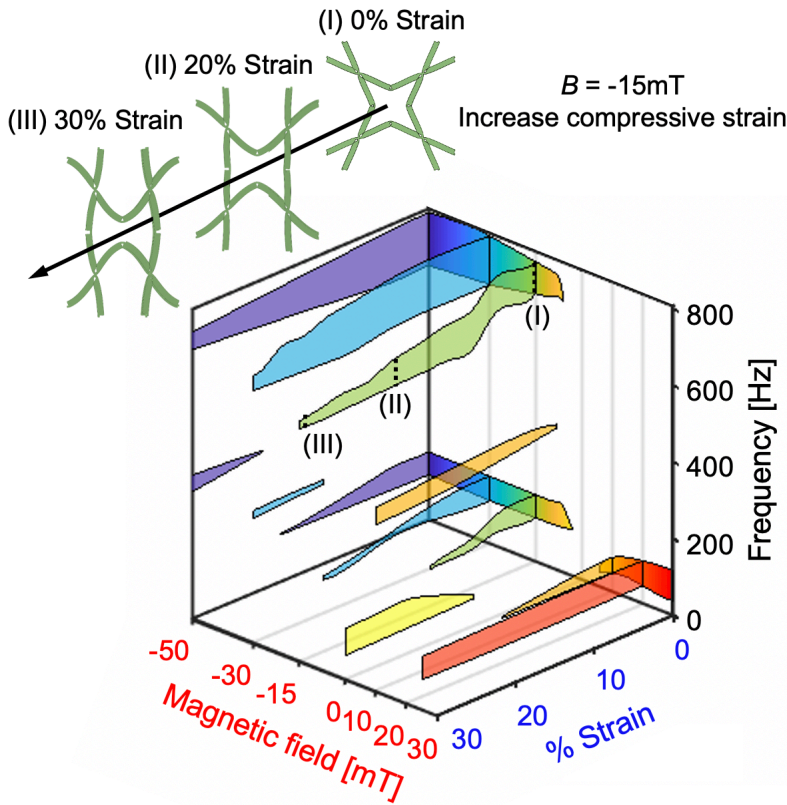

**Figure 5**. 3D plots of the bandgaps under different combinations of magnetic field and strain.

**Figure 5** shows the 3D plot of the bandgap widths under different combinations of the magnetic field and mechanical compression. The backplane at the zero strain shows the magnetic field-dependent bandgaps. Contours projecting from that plane represent states with a constant magnetic field but increasing strain. Each magnetic field state is represented by a single color for visualization. Small bandgaps (less than 20Hz) are considered insignificant and are omitted for clarity. In agreement with Figure 4a, in the unactuated state (zero magnetic field and zero strain) no bandgaps are observed. Actuation by either stimulus, however, can lead to the formation of bandgaps. This is in contrast to other acoustic metamaterials that can adjust the bandgaps under a certain stimulus but never exhibit an 'off' state where all wave propagation is permitted. This feature can be a useful feature for applications in waveguiding structures where the waveguide can be turned on and off.

When only a mechanical compressive load is applied to the metamaterial, a single bandgap starts to emerge at a compressive strain of ~10%, indicating the acoustic performance of the metamaterial array is relatively stable without the magnetic field. In the negative magnetic field direction (bending cases) bandgaps open quickly and then change slightly as the contact of gaps in the asymmetric joint restricts the shape change, which is similar to the evolution of the stiffness. As the compression is applied, as shown in Figure 4b, the vertical walls begin to transition to the outwards folding position. This causes the contact in those joints to be lost which introduces a shift in the bandgaps as shown in the -15mT (green) and -30mT (blue) contours. The -50mT case does not exhibit this transition because contact is maintained through 30% compression. For the positive magnetic field (folding) there is no contact for low magnetic field strengths. For this reason, the bandgaps tend to evolve more gradually. Similar to what is shown at 7mT in Figure 4c, for the 10mT case (light orange) the mechanical compression causes the low-frequency bandgap to close but opens a higher frequency gap. This is due to the transition from outwardly folding vertical walls to a much straighter position. Under 20mT, however, the magnetic field leads to a transformation of the metamaterial topology to a more stable one. The same deformation mode is maintained throughout compression, which causes the corresponding bandgap (red) to remain constant as the strain is increased.

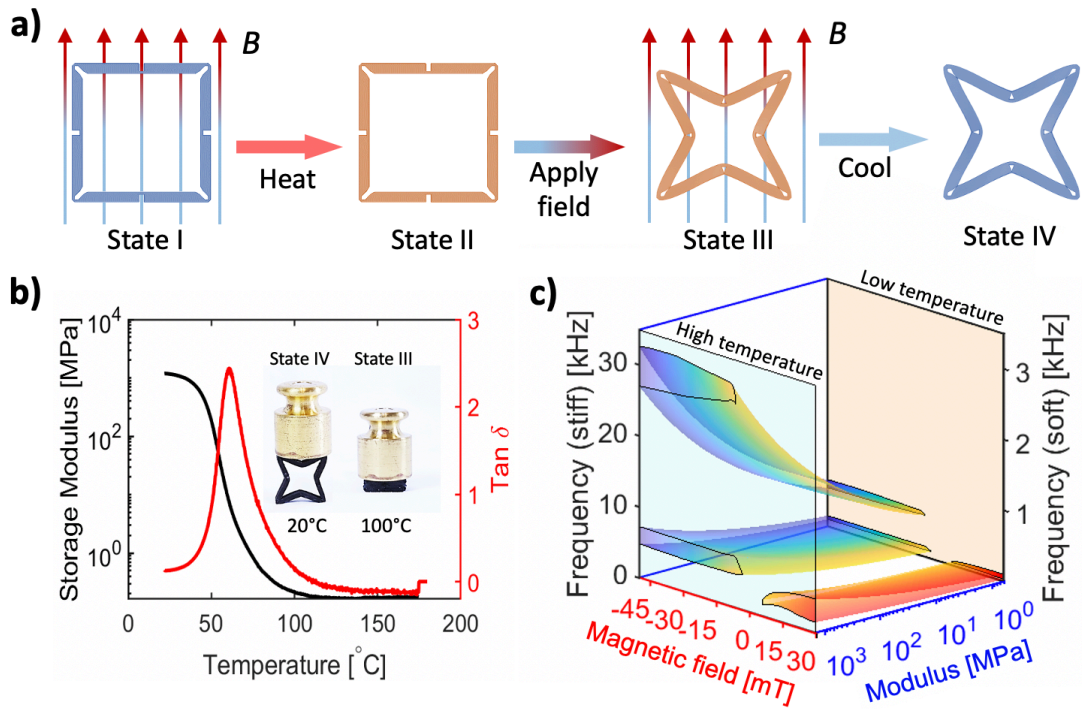

**Figure 6**. The global frequency tunability by using M-SMPs. a) The programming cycle for the metamaterial involves heating the metamaterial, applying the magnetic field, and cooling it to room temperature to lock the shape. b) The temperature-dependent material properties of the M-SMP system. c) The field and modulus-dependent bandgap locations.

**2.4 Global frequency tunability by magnetic shape memory polymers**

The metamaterials can obtain additional functionality by using M-SMPs (40), which have the advantage of shape locking due to the shape memory effect and tunable stiffness by temperature. The heating and actuation sequence to create the locked shape are shown in **Figure 6a**. At room temperature (State I) the unit cell cannot be actuated due to the high modulus of the M-SMP. As the M-SMP is heated (State II) it becomes soft and can be actuated into any of the previously demonstrated deformation modes (State III). Cooling the material back down to room temperature locks the final shape (State IV). When the shape is locked, the compressive forces and the magnetic field can be removed and the desired configuration can be retained. The extensive range of material properties obtainable by using the M-SMP is depicted in **Figure 6b**. The insets show a locked M-SMP unit cell at State IV supporting a 100g mass (~86 times of the unit cell weight) compared with the same cell at the high temperature (State II) supporting the same weight, demonstrating the drastic difference in properties.

The ability to control the modulus leads to another advantage of this metamaterial with widely tunable bandgaps. As shown in Figures 4 and 5, while the deformation mode branching using magneto-mechanical coupling allows for the control of the bandgaps within a frequency range that spans from ~40Hz to ~800Hz, which we term as local tunability, the large change in the material modulus allows for a much larger shift of all the bandgap frequencies to a higher frequency range, which we term as global tunability. This is demonstrated in **Figure 6c**. At the elevated temperature (well above $T_g$), the M-SMP has a rubbery modulus of 0.6MPa, which enables fast and large strain actuation. At room temperature, the M-SMP is in the glassy state with a Young's modulus of ~1.5GPa. The Young's modulus spans three orders of magnitude by altering the temperature, which can be harnessed to further tune the global bandgaps. The bandgaps of the metamaterial are observed in the 800Hz range at high temperatures (low modulus, back-plane of the plot in Figure 6c), but at low temperatures (high modulus, front-plane) the range of these bandgaps can extend up to over 32,000Hz. The lower bandgap that is around 110Hz wide at high temperature becomes over 5,500Hz wide at room temperature. This global and local bandgap shift makes this versatile metamaterial with deformation mode branching a new group of active metamaterials with great tunability.

**3. Conclusion**

We have presented a new magneto-mechanical metamaterial that allows great shape and property tunability through a novel concept of deformation mode branching. This new metamaterial uses an asymmetric joint design using hard-magnetic soft active materials that permits two distinct actuation modes (bending and folding) when opposite-direction magnetic fields are applied. The subsequent mechanical compression leads to the deformation mode branching where the metamaterial transforms into two distinctly different shapes with one exhibiting a topological transformation that has a significant reduction in the number of the pores in the metastructure. Such dramatic shape change also empowers great tunability in properties, such as mechanical stiffness and acoustic bandgaps. The unactuated metastructure has no bandgaps, which then emerge and vary under the coupled magneto-mechanical loading. This metamaterial design can also be incorporated with the magnetic shape memory polymers to allow global stiffness tunability, which further increases the achievable bandgap shift. The combination of magnetic and mechanical actuation as well as shape memory effects thus enable unmatched tunable properties and can lead to a new paradigm of metamaterials.

**Experimental Section**

*Hard-magnetic soft active material preparation*. For the hmSAM used in this study, a 20:1 by weight mixture of polymer to curing agent of Sylgard 184 Elastomer Kit (PDMS) (Fisher Scientific, Hampton, NH, USA) was used. A 20:1 ratio leads to a softer final material than the recommended 10:1 ratio. 15% vol of NdFeB particles (average particle size of 25μm, Magnequench, Singapore) were added to the PDMS resin mixture, which was then injected into polyvinyl alcohol (PVA) molds 3D printed by using a Creality Ender 5 FDM printer (Creality, Shenzhen, China). The filled molds were thoroughly degassed to ensure no trapped air bubbles. After degassing, the molds were covered by a glass slide and the hmSAMs were cured at 80°C for 2 hours in an oven. PVA is soft at high temperature so the cured hmSAM could be easily extracted.

*Metamaterial array fabrication*. Neighboring cells (after magnetization) were fused together in a checkerboard pattern using the same PDMS for the unit cell but without the magnetic particles.

*Magnetic shape memory polymer preparation*. The M-SMP used in this paper is an acrylate-based polymer modified from previous work (40). The SMP matrix contains aliphatic urethane diacrylate (Ebecryl 8413, Allnex, GA, USA), 2-Phenoxyethanol acrylate (Allnex), isobornyl acrylate (Sigma-Aldrich, St. Louis, MO, USA) mixed in a ratio of 10:30:60 by weight. The curing of the resin was enabled by the use of a thermal initiator (2,2'-Azobis(2-methylpropionitrile), 0.4 wt%). To prepare the M-SMP, 15% volume percent of NdFeB particles (size of 25μm) was added into the above resin matrix. Additionally, 2 wt% of fumed silica (Sigma-Aldrich) was mixed to enable proper dispersion of the magnetic particles. The process for fabricating a unit cell with this material was the same as that of the hmSAMs except the thermal curing process was done at 80°C for 4 hours with a post-curing step at 120°C for 0.5 hours.

*Unit cell magnetization*. A small, home-made coil with a high pulse magnetic field (~1.5T) was used to magnetize the embedded NdFeB particles, which retained their polarity after the field was removed due to the high magnetic coercivity. For the magnetization profile shown in Figure 1b, the hmSAM unit cell was first deformed into a "plus-sign" shape and held by a mold to retain the shape. The mold was placed inside the coil and then the impulse magnetic field was applied. For the M-SMP cells, they were mechanically programmed into the "plus-sign" shapes at high temperature, locked in place by shape memory effects, and then magnetized by the impulse magnetic field.

*Experimental setup.* For magnetic actuation, the unit cell or the metamaterial were placed between a pair of Helmholtz coils spaced 200mm apart. The coils were attached in parallel to a Preen ADG-L-160-25 4kW power supply (AC Power Corp., Taiwan, China). The current delivered to the coils was varied in order to control the magnetic field in between the coils. To apply a force during actuation, a 0.25mm nylon line was passed through 22AWG brass tubes that were glued to each corner of the unit cell using the same PDMS as above. Lines connected to the left side of the metamaterial were pulled through the center of the right coil and *vice versa*. On one side the lines were connected to a fixture. On the other side the lines were fed via pulleys to a universal material testing machine (Model 41, MTS, Eden Prairie, MN, USA). As the MTS probe was raised upwards, tension was applied to the lines, compressing the cell. A more detailed depiction of the test setup is included in Section S2 of the Supporting Information. The compression strain is defined as mechanical displacement during the mechanical loading divided by the length of the unit cell or array under the magnetic actuation (before applying mechanical loading).

## Acknowledgments

H.J.Q acknowledges the support of an Air Force Office of Scientific Research grant (AFOSR-FA9550-19-1-0151; Dr. B.-L. "Les" Lee, Program Manager). R.Z. acknowledges the support of the National Science Foundation (NSF) Career Award (CMMI-1943070), NSF Award (CMMI-1939543), and The Ohio State University Materials Research Seed Grant Program, an NSF-MRSEC grant (DMR-1420451).